

\font\twelverm=cmr10 scaled 1200    \font\twelvei=cmmi10 scaled 1200
\font\twelvesy=cmsy10 scaled 1200   \font\twelveex=cmex10 scaled 1200
\font\twelvebf=cmbx10 scaled 1200   \font\twelvesl=cmsl10 scaled 1200
\font\twelvett=cmtt10 scaled 1200   \font\twelveit=cmti10 scaled 1200
\font\twelvesc=cmcsc10 scaled 1200
\skewchar\twelvei='177   \skewchar\twelvesy='60
\def\twelvepoint{\normalbaselineskip=12.4pt
  \medskipamount=7.2pt plus2.4pt minus2.4pt
  \def\rm{\fam0\twelverm}          \def\it{\fam\itfam\twelveit}%
  \def\sl{\fam\slfam\twelvesl}     \def\bf{\fam\bffam\twelvebf}%
  \def\mit{\fam 1}                 \def\cal{\fam 2}%
  \def\tt{\twelvett}
  \def\sc{\twelvesc}
  \def\nullspace{\nulldelimiterspace=0pt \mathsurround=0pt }
  \def\big##1{{\hbox{$\left##1\vbox to 10.2pt{}\right.\nullspace$}}}
  \def\Big##1{{\hbox{$\left##1\vbox to 13.8pt{}\right.\nullspace$}}}
  \def\bigg##1{{\hbox{$\left##1\vbox to 17.4pt{}\right.\nullspace$}}}
  \def\Bigg##1{{\hbox{$\left##1\vbox to 21.0pt{}\right.\nullspace$}}}
  \textfont0=\twelverm   \scriptfont0=\tenrm   \scriptscriptfont0=\sevenrm
  \textfont1=\twelvei    \scriptfont1=\teni    \scriptscriptfont1=\seveni
  \textfont2=\twelvesy   \scriptfont2=\tensy   \scriptscriptfont2=\sevensy
  \textfont3=\twelveex   \scriptfont3=\twelveex  \scriptscriptfont3=\twelveex
  \textfont\itfam=\twelveit
  \textfont\slfam=\twelvesl
  \textfont\bffam=\twelvebf \scriptfont\bffam=\tenbf
  \scriptscriptfont\bffam=\sevenbf
  \normalbaselines\rm}

\def\beginlinemode{\endmode
  \begingroup\parskip=0pt \obeylines\def\\{\par}\def\endmode{\par\endgroup}}
\def\beginparmode{\endmode \begingroup \def\endmode{\par\endgroup}}
\let\endmode=\par
{\obeylines\gdef\
{}}
\def\singlespace{\baselineskip=\normalbaselineskip}
\def\oneandahalfspace{\baselineskip=\normalbaselineskip \multiply\baselineskip
     by 3 \divide\baselineskip by 2}
\def\doublespace{\baselineskip=\normalbaselineskip \multiply\baselineskip by 2}
\newcount\firstpageno \firstpageno=2
\footline={\ifnum\pageno<\firstpageno{\hfil}\else{\hfil\twelverm\folio\hfil}\fi}

\def\raggedcenter{\leftskip=4em plus 12em \rightskip=\leftskip
  \parindent=0pt \parfillskip=0pt \spaceskip=.3333em \xspaceskip=.5em
  \pretolerance=9999 \tolerance=9999 \hyphenpenalty=9999 \exhyphenpenalty=9999
}
\parskip=\medskipamount \twelvepoint \overfullrule=0pt
\def\title {\null\vskip 3pt plus 0.3fill \beginlinemode
   \doublespace \raggedcenter \bf}
\def\author{\vskip 3pt plus 0.3fill \beginparmode \raggedcenter \sc}
\def\affil{\vskip 3pt plus 0.1fill \beginlinemode
   \oneandahalfspace \raggedcenter \sl}
\def\abstract{\vskip 3pt plus 0.3fill \beginparmode
   \oneandahalfspace \narrower ABSTRACT:~~}
\def\endtitlepage{\vfill\eject\beginparmode}
\def\subhead#1{\vskip 0.25truein{\raggedcenter #1 \par} \nobreak
   \vskip 0.25truein\nobreak}
\def\references{\subhead{References}
   \frenchspacing \parindent=0pt \leftskip=0.8truecm \rightskip=0truecm
   \parskip=4pt plus 2pt \everypar{\hangindent=\parindent}}
\def\endreferences{\beginparmode}
\def\refstyleprd{
 \gdef\r##1{$^{##1}$}                                  
 \gdef\refis##1{\indent\hbox to 0pt{\hss##1.~~}}  
 \gdef\citerange##1##2##3{$^{\cite{##1}-\setbox0=\hbox{\cite{##2}}\cite{##3}}$}
 \gdef\journal##1, ##2, ##3, ##4.{{\sl##1} {\bf ##2}, ##4 (##3).} }
\def\cmp{\journal Comm. Math. Phys.}
\def\np{\journal Nucl. Phys.}
\def\pr{\journal Phys. Rev.}
\def\pl{\journal Phys. Lett.}
\def\prl{\journal Phys. Rev. Lett.}
\def\mpl{\journal Mod. Phys. Lett.}
\def\jmp{\journal J. Math. Phys.}
\def\endit{\endmode\vfill\supereject\end}

\def\ts{\textstyle}
\def\gtwid{\mathrel{\raise.3ex\hbox{$>$\kern-.75em\lower1ex\hbox{$\sim$}}}}
\def\ucsb{Department of Physics\\University of California\\
          Santa Barbara, CA 93106}
\def\p{\partial}
\def\D{{\cal D}}
\def\Xd{X^\dagger}
\def\Tr{\mathop{\rm Tr}}
\def\sss{\scriptscriptstyle}
\refstyleprd
\catcode`@=11
\newcount\r@fcount \r@fcount=0
\newcount\r@fcurr
\immediate\newwrite\reffile
\newif\ifr@ffile\r@ffilefalse
\def\w@rnwrite#1{\ifr@ffile\immediate\write\reffile{#1}\fi\message{#1}}
\def\writer@f#1>>{}
\def\referencefile{
  \r@ffiletrue\immediate\openout\reffile=\jobname.ref%
  \def\writer@f##1>>{\ifr@ffile\immediate\write\reffile%
    {\noexpand\refis{##1} = \csname r@fnum##1\endcsname = %
     \expandafter\expandafter\expandafter\strip@t\expandafter%
     \meaning\csname r@ftext\csname r@fnum##1\endcsname\endcsname}\fi}%
  \def\strip@t##1>>{}}

\def\citeall#1{\xdef#1##1{#1{\noexpand\cite{##1}}}}
\def\cite#1{\each@rg\citer@nge{#1}}	
\def\each@rg#1#2{{\let\thecsname=#1\expandafter\first@rg#2,\end,}}
\def\first@rg#1,{\thecsname{#1}\apply@rg}	
\def\apply@rg#1,{\ifx\end#1\let\next=\relax
\else,\thecsname{#1}\let\next=\apply@rg\fi\next}
\def\citer@nge#1{\citedor@nge#1-\end-}	
\def\citer@ngeat#1\end-{#1}
\def\citedor@nge#1-#2-{\ifx\end#2\r@featspace#1 
  \else\citel@@p{#1}{#2}\citer@ngeat\fi}	
\def\citel@@p#1#2{\ifnum#1>#2{\errmessage{Reference range #1-#2\space is bad.}%
    \errhelp{If you cite a series of references by the notation M-N, then M and
    N must be integers, and N must be greater than or equal to M.}}\else%
 {\count0=#1\count1=#2\advance\count1
by1\relax\expandafter\r@fcite\the\count0,%
  \loop\advance\count0 by1\relax
    \ifnum\count0<\count1,\expandafter\r@fcite\the\count0,%
  \repeat}\fi}
\def\r@featspace#1#2 {\r@fcite#1#2,}	
\def\r@fcite#1,{\ifuncit@d{#1}
    \newr@f{#1}%
    \expandafter\gdef\csname r@ftext\number\r@fcount\endcsname%
                     {\message{Reference #1 to be supplied.}%
                      \writer@f#1>>#1 to be supplied.\par}%
 \fi%
 \csname r@fnum#1\endcsname}
\def\ifuncit@d#1{\expandafter\ifx\csname r@fnum#1\endcsname\relax}%
\def\newr@f#1{\global\advance\r@fcount by1%
    \expandafter\xdef\csname r@fnum#1\endcsname{\number\r@fcount}}
\let\r@fis=\refis			
\def\refis#1#2#3\par{\ifuncit@d{#1}
   \newr@f{#1}%
   \w@rnwrite{Reference #1=\number\r@fcount\space is not cited up to now.}\fi%
  \expandafter\gdef\csname r@ftext\csname r@fnum#1\endcsname\endcsname%
  {\writer@f#1>>#2#3\par}}
\let\r@ferences=\references
\def\references{\r@ferences\def\endmode{\r@ferr\par\endgroup}}
\let\endr@ferences=\endreferences
\def\endreferences{\r@fcurr=0
  {\loop\ifnum\r@fcurr<\r@fcount
    \advance\r@fcurr by 1\relax\expandafter\r@fis\expandafter{\number\r@fcurr}%
    \csname r@ftext\number\r@fcurr\endcsname%
  \repeat}\gdef\r@ferr{}\endr@ferences}
\def\range#1#2#3{\citerange{#1}{#2}{#3}}
\let\r@fend=\endpaper\gdef\endpaper{\ifr@ffile
\immediate\write16{Cross References written on []\jobname.REF.}\fi\r@fend}
\catcode`@=12
\citeall\r

\singlespace
\rightline{hep-th/9206085}
\rightline{UCSBTH--92--23}
\doublespace

\title A NEW CONSTRUCTION OF THE PENNER MODEL

\author Mark Srednicki
\affil\ucsb

\abstract
The free energy of the Penner model is shown to be closely related to the
integral over the two diagonalizing unitary matrices of a complex rectangular
matrix.

\endtitlepage
\baselineskip=16pt

Matrix models\r{bipz78} can be solved if and only if the original integrals
over matrix elements can be reduced to integrals over
eigenvalues\r{iz80,mehta81}.  In these models, the ``angular'' integrals over
the diagonalizing unitary matrices result in a numerical factor which is
usually and properly ignored, since it contains no useful information and
can be absorbed into the normalization of the partition function.  Here we
point out a surprising fact: the numerical factor resulting from the angular
integrals for a complex matrix with $N'$ rows and $N$ columns is closely
related to the partition function for the Penner
model\range{penner86}{dv91,cdl91}{tan92},
which in turn is closely related to the moduli space of Riemann surfaces.
We will give a precise statement and proof of the relationship, but
unfortunately are unable to provide a physical or intuitive explanation of it.

Let us begin with a brief review of the Penner model.
It is defined by the partition function\r{penner86,dv91}
$$Z(t,N)=\xi(N)\int\D\phi\,e^{\,Nt\Tr[\phi+\log(1-\phi)]}\eqno(1)$$
where $\phi$ is a hermitian $N\times N$ matrix, $t$ is a positive real
parameter, $\D\phi$ indicates separate integration over the real and imaginary
parts of each matrix element of $\phi$,
and $\xi(N)$ is a normalization constant (which we will not be
interested in).
The free energy $F(t,N)=\log Z(t,n)$
of this model can be expanded as
$$F(t,N)=\sum_{g=0}^\infty N^{2-2g}\sum_{n=1}^\infty
t^{2-2g-n}\chi_{g,n}\eqno(2)$$
where $\chi_{g,n}$ is the Euler character of moduli space of Riemann surfaces
with genus $g$ and $n$ punctures.  Furthermore, this model possesses a double
scaling limit\range{gm90}{ds90}{bk90} in which we take
$N\to\infty$ and $t\to t_{\rm c}=-1$ with $\mu=(t_{\rm c}-t)N$ held
fixed\r{dv91}.  That $\mu\sim N^\gamma$ with $\gamma=1$ indicates that the
resulting conformal field theory has central charge $c=1$.

The integral in eq.(1) is divergent, and therefore needs some modifications in
order to be well defined.  Following ref.[\cite{dv91}], we first change matrix
variables to $M=1-\phi$ and then arbitrarily restrict the integration
to matrices $M$ whose eigenvalues are all positive.  We then have
$$\eqalignno{
Z(t,N)&=\xi(N)\,e^{N^2t}\int\D M_+\,(\det M)^{Nt}\,e^{-Nt\Tr M}&(3)\cr
      &=e^{N^2t}\int_0^\infty
        \prod_{i=1}^N\left(d\lambda_i\,\lambda_i^{Nt}
        \,e^{-Nt\lambda_i}\right)\Delta^2(\lambda)          &(4)\cr}$$
where $\Delta(\lambda)=\prod_{i<j}(\lambda_i-\lambda_j)$
is the Vandermonde determinant and $\xi(N)$
has been chosen to simplify the last line.
This form for $Z(t,N)$ is well defined and the expansion of $F(t,N)$
in powers of $N$ and $t$ yields the values of $\chi_{g,n}$.

Consider now an $N'\times N$ matrix $X$ with complex entries and no other
special properties.  Models of this type have been studied
before\r{amp91,mp92}, but the information we will extract from
them is new.  Without loss of generality we can choose $N'\ge N$ and define
a positive real parameter $t=(N'-N)/N$.  We will think of $t$ and $N$ as the
independent parameters, with $N'=(1+t)N$.  Define the integral
$$I\bigl(t,N;f(x)\bigr)=\int\D X\,f\bigl(\Tr\Xd X\bigr)\eqno(5)$$
where $\D X$ indicates separate integration over the real and imaginary
parts of each matrix element of $X$, and $f(x)$ is any function for
which the integral converges.  By making a $U(N')\times U(N)$
transformation on $X$, we can bring it to the form
$$X=\pmatrix{\lambda_1^{1/2}&\ldots&0\cr
\noalign{\medskip}
             \vdots&\ddots&\vdots\cr
\noalign{\medskip}
             0&\ldots&\lambda_{\sss N}^{1/2}\cr
\noalign{\smallskip}
             0&\ldots&0\cr
\noalign{\medskip}
             \vdots&\ddots&\vdots\cr
\noalign{\medskip}
             0&\ldots&0\cr}\eqno(6)$$
where the diagonal entries $\lambda_i^{1/2}$ are all real and positive.
We then have
$$I\bigl(t,N;f(x)\bigr)=\int_0^\infty
                        \prod_{i=1}^Nd\lambda_i\,J(\lambda)\,
                        f\bigl({\ts\sum_j\lambda_j}\bigr),\eqno(7)$$
where the jacobian $J(\lambda)$ was computed in ref.[\cite{amp91}]:
$$J(\lambda)=\Omega(t,N)\,\Delta^2(\lambda)\prod_{i=1}^N\lambda_i^{Nt}.
                                                                  \eqno(8)$$
The numerical factor $\Omega(t,N)$, which was not computed in
ref.[\cite{amp91}], results from the integral over the two diagonalizing
unitary matrices.  $\Omega(t,N)$ will be our focus of attention.

Let us now choose $f(x)=e^{-Ntx}$.  This turns $I$ into
a product of $2N'N$ independent gaussian integrals, yielding the simple formula
$$I\bigl(t,N;e^{-Ntx}\bigr)=\left({\pi\over Nt}\right)^{N^2(1+t)}.\eqno(9)$$
On the other hand, putting $f(x)=e^{-Ntx}$ into eq.(7) and using eq.(8) gives
$$\eqalignno{
I\bigl(t,N;e^{-Ntx}\bigr)&=\Omega(t,N)\int_0^\infty
                           \prod_{i=1}^N\left(d\lambda_i\,\lambda_i^{Nt}
                           e^{-Nt\lambda_i}\right)\Delta^2(\lambda)&(10)\cr
\noalign{\medskip}
                         &=\Omega(t,N)\,e^{-N^2t}Z(t,N),&(11)\cr}$$
where $Z(t,N)$ is the partition function of the Penner model, as defined
in eq.(4).  Equation the right-hand sides of eqs.(9) and (11), we find the
bizarre result that $\Omega(t,N)$ is closely
related to the free energy of the Penner model:
$$\log\Omega(t,N)=-F(t,N)-N^2(1+t)\log(Nt/\pi)+N^2t\;.\eqno(12)$$
If we now analytically continue this formula to negative $t$, we can take
the double scaling limit $N\to\infty$ and $t\to -1$ with $\mu=-(1+t)N$
fixed.  In this limit, $F$ is not well defined, but its third derivative with
respect to $\mu$ is\r{dv91}.  At fixed $N$ we have
$$\bigl[\log\Omega\bigr]'''=-F'''-{N(3N+\mu)\over(N+\mu)^3}\eqno(13)$$
where a prime denotes $\p/\p\mu$.  In the $N\to\infty$ limit, $F'''$
becomes $[\mu\psi(\mu)]''$, where $\psi(\mu)$ is the Euler psi
function\r{dv91}, while the second term on the right hand side vanishes.

This establishes an unexpected connection between integrals over unitary
matrices and the Penner model.  At present there is no obvious reason why
this connection should exist, but it may be worthy of further investigation.

I would like to thank S.~Chaudhuri and J.~Distler for discussions,
and the Aspen Center for Physics, where the discussions took place.
This work was supported in part by NSF Grant No.~PHY-91-16964.

\references
\baselineskip=16pt

\refis{amp91}A. Anderson, R. C. Myers, and V. Periwal, \pl, B254,
1991, 89.\llap; \np, B360, 1991, 463.

\refis{bipz78}E. Brezin, C. Itzykson, G. Parisi, and J. B. Zuber,
\cmp, 59, 1978, 35.

\refis{cdl91}S. Chaudhuri, H. Dykstra, and J. Lykken, \mpl, A6, 1991, 1665.

\refis{dv91}J. Distler and C. Vafa, \mpl, A6, 1991, 259.

\refis{ds90}M. Douglas and S. Shenker, \np, B335, 1990, 635.

\refis{gm90}D. Gross and A. A. Migdal, \prl, 64, 1990, 717.

\refis{iz80}C. Itzykson and J. B. Zuber, \jmp, 21, 1980, 411.

\refis{bk90}E. Brezin and V. Kazakov, \pl, 236B, 1990, 144.

\refis{mehta81}M. L. Mehta, \journal Comm. Math. Phys., 79, 1981, 327.

\refis{mp92}R. C. Myers and V. Periwal, McGill Univ. preprint MCGILL--92--01
(January, 1992).

\refis{penner86}R. C. Penner, {\sl Perturbative series and the moduli space
of Riemann surfaces}, U.C. San Diego preprint, 1986.

\refis{tan92}Chung-I Tan, \pr, D45, 1992, 2862.

\endreferences\endit\end